# Screened Strong Coupling of Excitons in Multilayer WS$_2$ with Collective Plasmonic Resonances

Shaojun Wang,[†*] Quynh Le-Van,[†] Fabio Vaianella,[#] Bjorn Maes,[#] Simone Eizagirre Barker,[†] Rasmus H. Godiksen,[‡] Alberto G. Curto,[‡] and Jaime Gomez Rivas[†,‡*]

[†] Dutch Institute for Fundamental Energy Research, Eindhoven, The Netherlands

[#] Micro- and Nanophotonic Materials Group, Faculty of Science, University of Mons, 20 place du Parc, B-7000 Mons, Belgium

[‡] Department of Applied Physics and Institute for Photonic Integration, Eindhoven University of Technology, Eindhoven, The Netherlands

**ABSTRACT**: We demonstrate the strong coupling of direct transition excitons in tungsten disulfide (WS$_2$) with collective plasmonic resonances at room temperature. We use open plasmonic cavities formed by periodic arrays of metallic nanoparticles. We show clear anti-crossings with monolayer, bilayer and thicker multilayer WS$_2$ on top of the nanoparticle array. The Rabi energy of such hybrid system varies from 50 to 110 meV from monolayer to sixteen layers, while it does not scale with the square root of the number of layers as expected for collective strong coupling. We prove that out-of-plane coupling components can be disregarded since the normal field is screened due to the high refractive index contrast of the dielectric layers. Even though the in-plane dipole moments of the excitons decrease beyond monolayers, the strong in-plane field distributed in the flake can still enhance the coupling strength with multilayers. However, the screened out-of-plane field leads to the saturation of the Rabi energy. The achieved coherent coupling of TMD multilayers with open cavities could be exploited for manipulating the dynamics and transport of excitons in 2D semiconductors and developing ultrafast valley/spintronic devices.

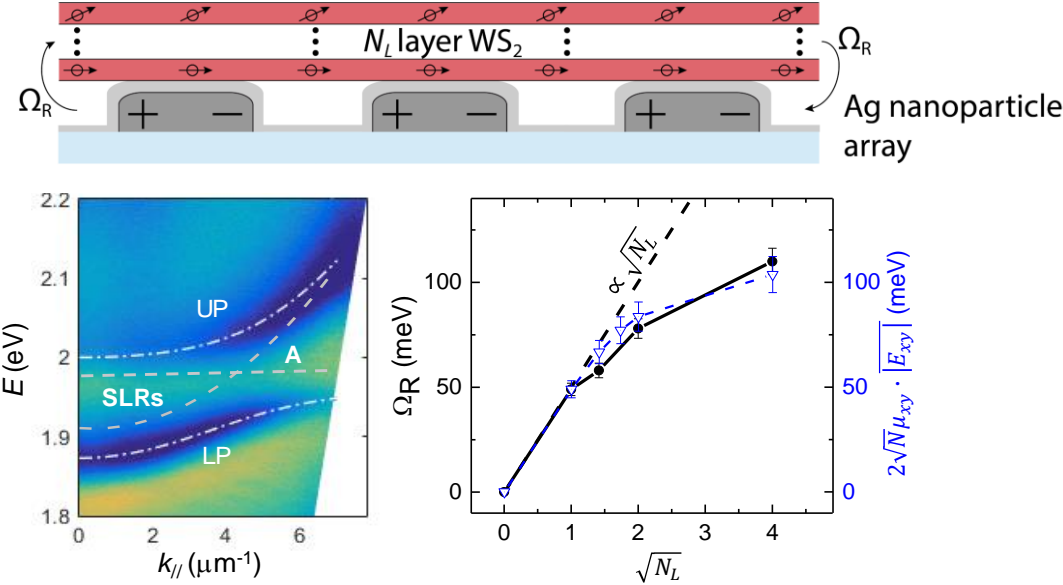

TOC Graphic

Atomically-thin transition metal dichalcogenides (TMDs, with chemical formula $MX_2$, M = Mo, W; X = S, Se, Te) are being intensively studied in the context of light-matter interaction after the indirect to direct band gap crossover in monolayers was established.[1–5] The remarkable optical response in such two-dimensional (2D) semiconductors originates from the strongly allowed excitonic resonances, displaying a large absorption coefficient in the visible and near-infrared.[1,6] The optically generated electron-hole pairs (excitons) in such 2D systems have exceptionally large binding energies (> 0.3 eV),[7] small Bohr radii (~ 1 nm),[8] and a new valley degree of freedom arising from the strong spin-orbit coupling and the absence of inversion symmetry of the monolayers.[9–11] These promising optoelectronic properties of excitons in 2D-TMDs provide an elegant platform to investigate strong light-matter interaction[12–16] as well as open possibilities for novel device architectures.

The regime of strong coupling between light and excitons in semiconductors to form exciton-polaritons is reached when the energy exchange between the excitonic transitions and a resonant optical mode is faster than their dephasing rates.[17,18] The spectral signature of strong coupling is the splitting of the absorption band of excitons into two new bands, corresponding to the upper and lower polaritonic states. The energy splitting of these states at resonance measures the coupling strength (*i.e.*, the Rabi energy, $\Omega_R$) and it depends on the electromagnetic field per photon ***E*** of the optical mode, the transition dipole moment ***μ*** of the exciton and $\sqrt{N}$, where $N$ is the number of excitons within the volume of the optical mode (*i.e.*, $\Omega_R = 2\sqrt{N}\boldsymbol{E}\cdot\boldsymbol{\mu}$).[17–19] Excitons in TMD monolayers with a large transition dipole moment have been demonstrated to couple strongly with different optical resonators, such as microcavities,[12,13,15,20] photonic crystals,[21] and plasmonic systems.[14,15,22–27] The hybridization of excitons in TMD monolayers with optical resonators plays an important role in fascinating phenomena, such as valley-selective chiral coupling,[22] thermally or electrically tuneable composition of the hybrid states,[23–25] and enhanced nonlinear optical susceptibility.[27] So far, most research has focused on the strong coupling of bright excitons in monolayers due to their well-known direct optical band gap and in-plane dipole moment.[12-15,20-27] For multilayer TMDs, the orientation of the excitonic dipole moments is still an open question and the energetically indirect transitions shown in their emission spectra are reminiscent of the ultrafast dephasing rate of direct transition excitons. Only two publications have recently reported strong coupling in different thickness of multilayer TMDs to plasmonic modes.[16,28] In both publications the coupling to single metallic nanoparticles has been investigated, where the mode volume is in the nanometric scale and delocalized properties arising from strong coupling, such as enhanced exciton-polariton transport[29,30] and long distance energy transfer,[31,32] cannot be achieved. Strong coupling of multilayer TMDs to extended plasmonic cavities is still an unexploited territory.

In this Letter, we demonstrate the strong coupling of excitons in $WS_2$ from monolayers to sixteen layers with optical modes in extended open plasmonic cavities formed by periodic arrays of metallic nanoparticles. The nanoparticles support localized surface plasmon resonances (LSPRs), i.e., coherent oscillations of the free electrons in the nanoparticle driven by the electromagnetic field. LSPRs are interesting for optical control of excitons in close proximity of the nanoparticles because their small electromagnetic mode volume can lead to high coupling efficiencies.[16,25,28,33,34] However, inherent to this small mode volume are also high losses.[33,35] Interestingly, periodic arrays of metallic nanoparticles support surface lattice resonances (SLRs), which are collective plasmonic resonances arising from the enhanced radiative coupling of LSPRs through diffracted orders in the plane of the array, also known as Rayleigh anomalies (RAs).[36–40] The quality factor of SLRs, which is inversely proportional to the loss, can be orders of magnitude larger than that of LSPRs.[36,37,40–43] Consequently, from the dynamic point of view, SLRs increase the cycles of Rabi oscillations in the strongly coupled system before dephasing. In addition, the dispersive behavior of SLRs, mainly defined by the periodicity of the array and the polarizability of the constituent nanoparticles, provides an easy mechanism to observe the evidence of strong coupling, i.e., anti-crossing of lower and upper polariton bands.[35]

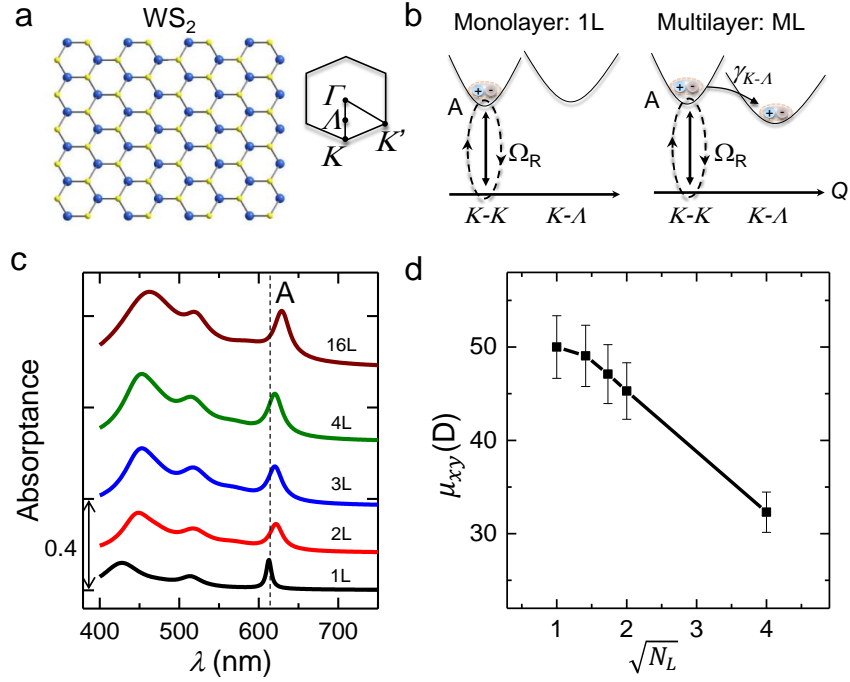

**Figure 1.** *Absorption spectra and energy diagram of $WS_2$.* (a) Crystal structure (top view) of $WS_2$ and its associated Brillouin zone. (b) Schematic energy diagram of direct excitonic transitions ($K$-$K$ or $K'$-$K'$) in $WS_2$ monolayer and multilayer, strongly coupled to an optical resonator as represented by the Rabi oscillation $\Omega_R$. For the multilayer, the excitons generated at the $K$ / $K'$ points of the Brillouin zone are thermally scattered into the energetically lower $\Lambda$ valleys ($K$-$\Lambda$). $Q$ represents the center of mass momentum of the exciton. (c) Absorption spectra of a monolayer (1L, black line), bilayer (2L, red), trilayer (3L, blue), four-layer (4L, green) and sixteen-layer (16L, maroon) of $WS_2$ on polydimethylsiloxane (PDMS) substrates, calculated by fitting the experimental transmittance spectra using a multi-Lorentzian model and the transfer matrix method. The peaks at lowest energy correspond to A-exciton transitions in $WS_2$. (d) In-plane dipole moments of the A excitons of $WS_2$ as a function of the square root of the number of $WS_2$ layers, $\sqrt{N_L}$.

We used atomically thin layers of $WS_2$ (Figure 1a) to reach the strong coupling regime due to the sharp and intense absorption peak of their A excitons.[15,20] To characterize the in-plane excitonic dipole moments, we have measured the normal incidence transmittance spectra through two $WS_2$ flakes with a monolayer (abbreviated to 1L) and a multilayer (ML) formed by 16 layers (16L) on sample I (Figure 2a) and another 1L, bilayer (2L), trilayer (3L), and four-layer (4L) on sample II (Figure S3a). The in-plane refractive index (Figures S4c,d) and the normal incidence absorptance spectra (Figure 1c) of each region were calculated by fitting the transmittance spectra using a multi-Lorentzian model and the transfer matrix method (TMM) (Figures S4a and b).[15,44] As can be appreciated in Figure 1c and Figure S5a, the absorption peak around 613 nm of A excitons in the 1L is redshifted by 9 nm in 2L, 3L and 4L measured on the same flake. When the thickness of the flake is increased to 16L, the A exciton resonance is further shifted towards 629 nm, approaching the bulk value. In addition, the absorptance per layer at the A excitonic transition (Figure S5b) decreases by increasing the number of layers. The linewidth ($\gamma_A$) of the ML is nearly constant at ~ 60 meV and broader than the linewidth of 1L ~ 25 meV. To evaluate the in-plane dipole moment $\mu_{xy}$, the absorptance per layer ($abs$) is spectrally integrated for the A excitonic transition (Figure S5a). We have determined the evolution of $\mu_{xy}$ as a function of the number of layers based on the relationship $\mu_{xy} \propto \sqrt{abs}$ and using the value for 1L of $\mu_{xy}$ ~ 50 Debye (D) obtained from the literatures.[34,45] The results (Figure 1d) reveal that the in-plane dipole moments gradually decrease to ~ 30 D for 16L, which we attribute to the increase of out-of-plane dipole moments when the interlayer coupling delocalizes excitons between planes.[16] Understanding the excitonic dipole components in WS$_2$ will help us to elucidate the nature of the coupling with the electric field of the SLRs as addressed below.

The absorption peak of A excitons in the 1L and ML is known to arise from direct band gap transitions at the *K* and *K'* points of the associated first Brillouin zone (Figure 1a, inset).[1,2] From the photoluminescence (PL) spectra (Figure S3d), excitons generated at the *K* / *K'* in the ML are mostly thermally scattered into the energetically lower *Λ* valleys (*K-Λ*).[46] The increased dephasing paths in ML homogeneously broaden the linewidth of A excitonic transitions in absorption and PL spectra. The energy diagrams in the exciton picture are shown in Figure 1b for the case of the 1L and the ML. Both 1L and ML can strongly couple with a resonant optical mode when their direct transitions exchange energy with this mode at a Rabi rate ($\Omega_R$) faster than the intravalley dephasing rate ($\gamma_{K-K}$) or valley scattering dephasing rate ($\gamma_{K-\Lambda}$) of the A excitons (i.e., $\Omega_R > \gamma_{K-K}, \gamma_{K-\Lambda}$). A similar strong coupling mechanism has been proved in a push-pull molecular system with large Stoke shift emission.[47]

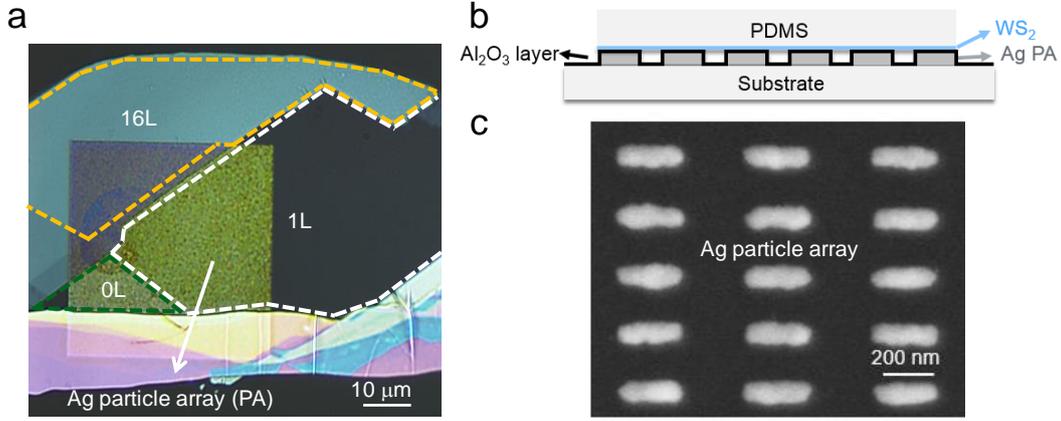

**Figure 2.** *Sample images.* (a) Optical micrograph (top view) of a $WS_2$ flake deposited on the Ag particle array (square of side 50 μm). The bare particle array (0L), 1L and 16L regions are encompassed by the green, white, and orange dashed curves, respectively. (b) Schematic side view of the sample. $WS_2$ on superstrate of PDMS is spatially separated from the particle array by a conformal alumina layer. (c) Scanning electron microscope image (top view) of the particle array.

We have coupled the 2D-semiconductor to SLRs confined on plasmonic lattices, as the one shown in the scanning electron microscope image of Figure 2c. The plasmonic lattice consists of an array of silver (Ag) nanoparticles with a horizontal lattice constant $P_x = 420$ nm and a vertical lattice constant $P_y = 200$ nm. The individual nanoparticles are nanorods with a height of 35 nm, a length of 230 nm, and a width of 70 nm. We have chosen a rectangular shape for the nanoparticles to increase their polarizability. This polarizability is proportional to the particle volume and it can be increased by making the particles longer while keeping the width constant to avoid shifting the resonance frequency for a polarization along the width. Increasing the polarizability of the nanoparticles enhances the extinction and facilitates the excitation of SLRs. The array, with a size of $50 \times 50$ μm$^2$, was fabricated on a fused silica substrate by electron-beam lithography. Each nanoparticle is coated with a 3-nm-thick atomic-layer-deposited Al$_2$O$_3$ film that protects it from degradation and also prevents charge transfer between the semiconductor and silver. The investigated sample consists of a mechanically exfoliated WS$_2$ flake covering a plasmonic nanoparticle array (Figure 2a). To ensure high-quality samples, we exfoliated flakes on a flexible polydimethylsiloxane (PDMS) substrate (with a thickness of 100 μm) and we softly transferred them onto a particle array. The PDMS substrate was left as a dielectric superstrate coating for the particle array (see details in **Methods**). The final sample geometry is shown in Figure 2b. The regions without flake (0L) and with 1L and 16L of WS$_2$ are indicated in Figure 2a by the green, white and orange curves, respectively. They show a significant change in color depending on whether the flake is on the fused silica substrate or on the particle array. The color change indicates that the optical resonances supported by the nanoparticle array can significantly modify the reflectivity of the atomically thin material. We should note that the color change on the particle array is not simply due to the thin air spacer between the substrate and the PDMS, since the flake still shows a similar color change after the superstrate of PDMS on top is peeled off (Figure S2c).

In order to obtain the dispersion of the coupled system, we have recorded angle-resolved white-light reflection spectra of sample I with a Fourier spectroscopy microscope.[15] The orientation of the plasmonic lattice is fixed by aligning its long axis vertical to the slit of the spectrometer. The direction of a polarizer parallel or perpendicular to the slit defines the *p*-(TM) or *s*-(TE) polarization of the reflected light, respectively. The angular dispersion of SLRs along the (0, ±1) RAs of the bare particle array measured with *p*-polarized light is shown in Figure 3a. We plot the evolution of the spectra with varying angle of incidence $\theta = 0°$ to $35°$ in Figure 3d. The angular dispersion of SLRs has a parabolic shape (Figures 3a and S7a) and results from the radiative coupling between the LSPRs of the individual particles at 525 nm (2.36 eV), enhanced by the in-plane diffraction of the RAs. We have extracted the linewidths of the SLRs by fitting the reflection spectra to a Fano-like line shape.[48] The spectral full width at half maximum (FWHM) of SLRs, $\Gamma_{SLRs}$, also has a dispersive behavior (Figure S7d), increasing from 43 meV at $\theta = 0°$ to 73 meV at $\theta = 35°$ as the weight fraction of LSPRs increases in the hybrid SLRs.

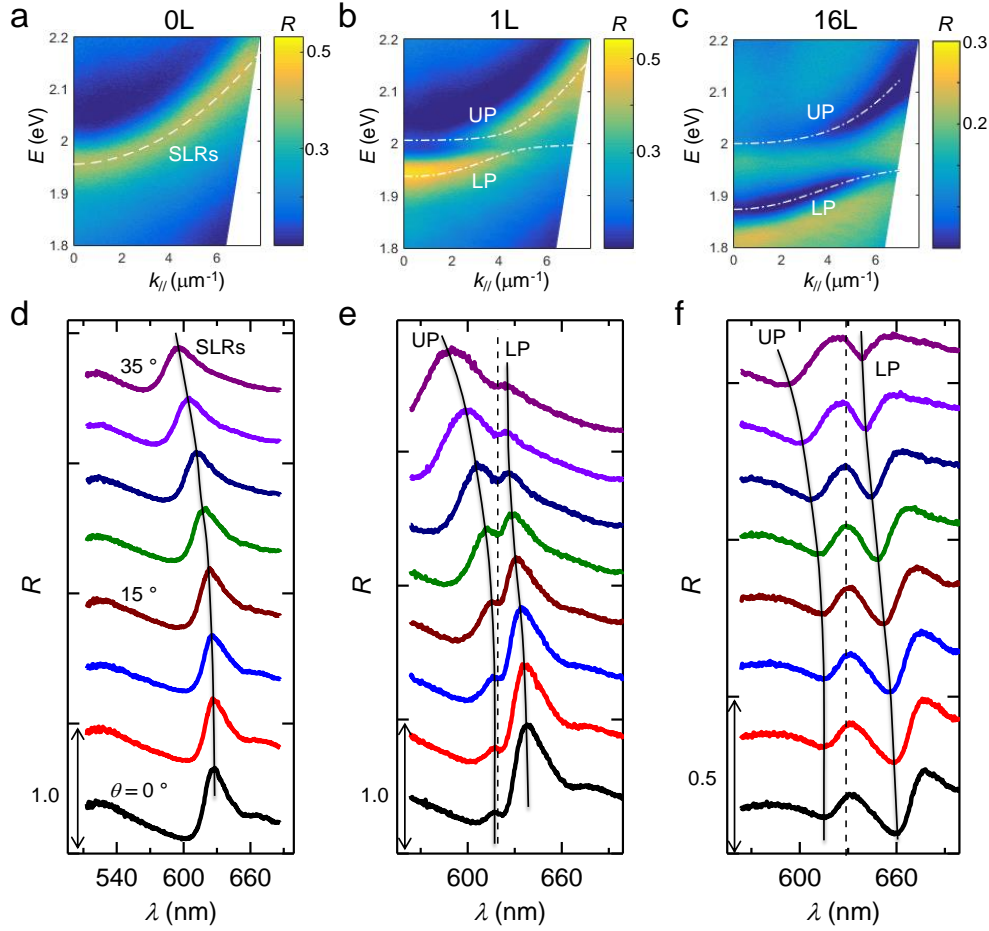

**Figure 3.** *Dispersion measurements*. (a) Angle-resolved reflection spectra of the bare nanoparticle array (0L), (b) 1L WS$_2$ on top of the array, and (c) 16L on top of the array. The spectra are analyzed in *p*-(TM) polarization. The angular dispersion of each sample is fitted with the two coupled oscillator model. The dashed white curves in (a) describe the dispersion of the surface lattice resonances (SLRs) of the bare particle array. The dashed-dotted white curves in (b-c) show the dispersion of the upper polariton (UP) band and the lower polariton (LP) band. (d-f) The reflection spectra as a function of the angle of incidence from $\theta = 0°$ to 35° obtained by cutting through the corresponding Fourier images in (a-c). The solid lines in (d-f) are guides to the eye for the angular dispersion of SLRs and polariton bands. The vertical dashed lines in (e) and (f) represent the A excitonic energy in WS$_2$ obtained from the absorption (Fig. S5a) and reflection (Fig. S6) spectra.

On the part of the monolayer transferred on top of the array, the parabolic dispersion curves split into two new bands corresponding to the upper (UP) and lower polariton (LP) bands (Figure 3b). This unambiguous anti-crossing is also seen in the angular evolution of spectra shown in Figure 3e. The hybrid system can be approximately described with a two coupled harmonic oscillator model:[39,40,49]

$$\begin{bmatrix} \omega_{SLRs} - i\frac{\Gamma_{SLRs}}{2} & g \\ g & \omega_A - i\frac{\gamma_A}{2} \end{bmatrix} \begin{pmatrix} \alpha \\ \beta \end{pmatrix} = \omega \begin{pmatrix} \alpha \\ \beta \end{pmatrix}, \quad (1)$$

where $\omega_{SLRs}$ and $\omega_A$ are the energies of the bare SLRs and A excitons respectively, $\gamma_A$ represents the linewidth of excitons by considering the linewidth as the dephasing rate of excitons due to homogeneous broadening, and $g$ is the coherent coupling strength. Diagonalizing the Hamitonian matrix yields new polaritonic eigenvalues $\omega_\pm$, defining the energies of UP and LP bands, and the Hopfield coefficients $\alpha$ and $\beta$,[50] the square of which defines the weight fractions of excitons and SLRs with $|\alpha|^2 + |\beta|^2 = 1$. By fitting the peak positions of the UP and LP bands in Figures 3b and e to the model, we have extracted $\omega_{SLRs}$ and the weight fractions assuming $\omega_A = 2.006$ eV, which was determined from the absorption peak of A exciton. The hybrid bands and the weight fractions of the LP are shown in Figures S7b and e, respectively. The Rabi energy, $\Omega_R = 2g = 47$ meV, is given at the in-plane wave vector $k_{//} = 3.5$ μm$^{-1}$ when the UP or LP state is half mixed with SLRs and half with excitons, *i.e.*, $|\alpha|^2 = |\beta|^2 = \frac{1}{2}$. This analysis indicates that the spectra of uncoupled SLRs redshift 4 nm (Figures S7a and b) when the 1L WS$_2$ is transferred on top of the particle array.

Using the same method, we have characterized the angular dispersion and coupling parameters in all the regions of interest on Samples I and II. The reflection spectra of 16L on the particle array of sample I, and 1L, 2L, 4L on sample II are plotted in Figures 3c and f, and Figure S8, respectively. As illustrated in these figures with the solid curves, the LP and UP bands are obtained by fitting the peaks for the 1L, 2L, 3L and 4L. The asymmetric Fano line shape of SLRs, due to the interference between LSPRs and RAs, gives rise to a better visibility of the dips for the 16L. Therefore, for this sample we fit the dips instead of the peaks. We note that this specific Fano-type resonance results in the simultaneous observation of both peak and dip splitting, as also recently reported in terms of the *s*-(TE) polarized mode by Liu et al.[26]

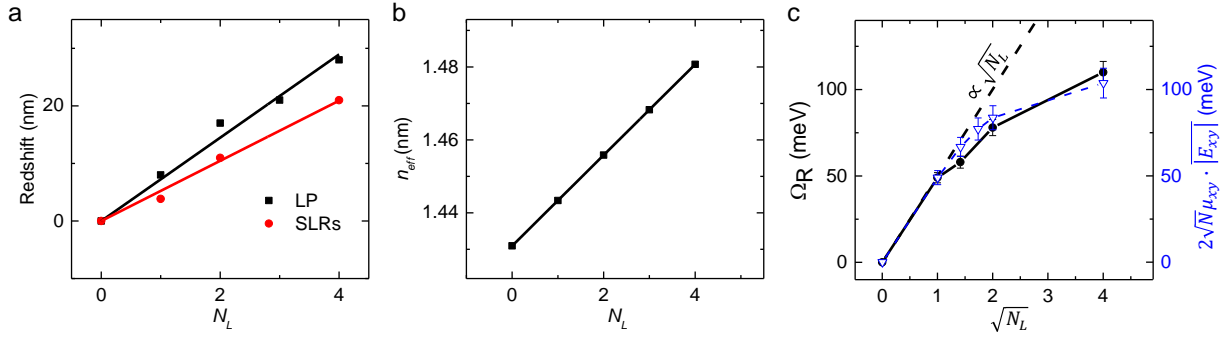

**Figure 4.** *Interaction between SLRs and WS$_2$.* (a) Linear redshift of the LP band (black squares) and the SLRs (red dots) as a function of the number of layers ($N_L$ from 0L to 4L) of WS$_2$. The peak wavelengths of LP band are obtained from the consistent reflection and transmission spectra (Fig. S10) measured at normal incidence. The resonant wavelengths of bare SLRs are extracted from fitting the angular dispersion of the coupled system (Fig. S9). (b) Effective refractive index of the particle array as a function of $N_L$ calculated from the grating equation and the linear slope of bare SLRs shown in (a). (c) Measured (black circles) and estimated (blue triangles) Rabi energy $\Omega_R$ as a function of the square root of $N_L$. The estimation is based on the formula of vacuum coupling strength $g = \sqrt{N}\bm{E}\cdot\bm{\mu}$, and herein we only account for the in-plane coupling due to the screened normal field. The dashed black line represents the linear relationship between $\Omega_R$ and $\sqrt{N_L}$ when the dot product of $\bm{\mu}$ and $\bm{E}$ is constant.

The energy splitting revealed by the two solid lines in Figures 3c and S8 becomes larger at the resonant condition (around $\theta = 20°$) when increasing the thickness of the flake. In addition, the LP band at normal incidence redshifts linearly as function of number of layers (Figure 4a) when comparing the reflection and extinction spectra of 0L to 4L (Figure S10) on the particle array. In order to extract the bare SLRs, we fitted the angular dispersions of different regions with the coupled oscillator model (Figures S9) and found that the bare SLRs also varies linearly (Figure 4a) from 0L to 4L. From the grating equation (see **Methods**), the resonant wavelength of (0, ±1) RAs can be calculated by $\lambda_{RAs}= n_{eff}\cdot P_x$, where $n_{eff}$ is the effective refractive index defining the phase velocity of the RAs. In the case of the bare particle array, $\lambda_{RAs} \sim 601$ nm and we get $n_{eff} = 1.431$. Considering the same slope of RAs and SLRs, we obtain the $n_{eff}$ of 1L ($n_{eff}$

= 1.443) to 4L ($n_{eff}$ = 1.481) and a variation of $n_{eff}$ per layer of 0.012. Such large change of effective refractive index by just adding an atomic layer on top of the particle array highlights the strong interaction between the field around the particle array and the high background refractive index ($n$ = 2.76, see Figure S4c) of $WS_2$. In addition, the linear response of $n_{eff}$ in Figure 4b suggests that the interacting field amplitude in 0L and 4L can be considered as a constant.

Table 1 Coupling parameters and evaluation of the strong coupling regimes.

|  | $k_{//}$ (μm$^{-1}$) | $\Gamma_{SLRs}$ (meV) | $\gamma_A$ (meV) | $\Omega_R$ (meV) | $\Omega_R^2 > (\Gamma_{SLRs}^2 + \gamma_A^2)/2$ | $\Omega_R > \Gamma_{SLRs}, \gamma_A$ |
|---|---|---|---|---|---|---|
| **1L**[a] | 3.5$^I$, 4.4$^{II}$ | 54$^I$, 60$^{II}$ | 28$^I$, 25$^{II}$ | 47$^I$, 52$^{II}$ | yes | no |
| **2L** | 3.7 | 55 | 59 | 58 | yes | no |
| **4L** | 4.5 | 61 | 62 | 78 | yes | yes |
| **16L** | 4.2 | 59 | 60 | 110 | yes | yes |

[a] Two different 1L samples (I and II) on the particle array were characterized.

We continue the analysis with a comparison of the coupling strength between the SLRs and the excitons in $WS_2$ flakes of different thicknesses. We summarize the key coupling parameters in Table 1, including the resonant in-plane wave-vector $k_{//}$, the corresponding $\Gamma_{SLRs}$, $\gamma_A$, $\Omega_R$ and the evaluation of the strong coupling regime. Table 1 shows that the higher quality of 1L on sample II (with bare absorptance of 13% at $\omega_A$ = 2.023 eV, Figure S5a) yields a slightly higher coupling strength compared to 1L on sample I. From the $\Omega_R$ in table 1, we can clearly see that the coupling strength is enhanced by increasing the number of layers from 1L to 16L. However, $\Omega_R$ as a function of $\sqrt{N_L}$ (black dots in Figure 4c) in the case of ML is significantly lower than the value expected from the formula of collective strong coupling $\Omega_R \propto \sqrt{N_L}$,[51] as noted by the dashed line in Figure 4c. This reveals that the scalar multiplication of **E** and **μ** as a function of $\sqrt{N_L}$ must decrease.

In order to claim strong coupling, we have to consider both the anti-crossing and the dynamics of the coupled system ruled by the competition between coherent Rabi-oscillations and various dephasing processes of the coupled system. The weak criterion of strong coupling is $\Omega_R^2 > (\gamma_A^2 + \Gamma_{SLRs}^2)/2$, namely, the splitting energy needs to be larger than the averaged damping rate of the bare oscillators in order to guarantee observation of two new peaks.[35] A stricter criterion is that the Rabi energy needs to overcome all the damping rates of the individual oscillators, i.e, $\Omega_R > \gamma_A$, $\Gamma_{SLRs}$. We evaluate both criteria and summarize them in the last two columns of Table 1. The coupling between the nanoparticle arrays with 1L to 16L $WS_2$ can reach the strong coupling regime under the weak criterion. For satisfying the strong criterion, we need to overcome the broad linewidth of SLRs for the case of 1L and enhance the coupling strength further for 2L.

To fully understand the coupling efficiency of SLRs with the 1L and the ML, we have simulated the anisotropic field enhancement distribution on a bare particle array and with a flake on the array using the finite-element method (details in **Methods**). The same parameters of the experimental geometry were used for the simulations, obtaining a similar bare SLR angular dispersion to the experiments (Figure 5a). Note that the simulated dispersion of bare SLRs is plotted herein as a function of in-plane wave vector distributed in PDMS instead of air during the measurements. When the 1L $WS_2$ is placed on the particle array (Figure 5b), the original bare SLR band displays a Rabi energy of 43 meV, a bit smaller than the experimental value (52 meV) due to the sharper simulated SLRs ($\Gamma_{SLRs}$ = 36 meV). The electric field intensity enhancement factor of the SLRs compared to free space at the energy of A excitons was calculated in terms of in-plane ($|E_{xy}|^2/|E_0|^2$) and out-of-plane components ($|E_z|^2/|E_0|^2$), respectively. In these calculations we consider the background refractive index of $WS_2$. Due to the typical Fano-type resonance of SLRs, we scan the angle of incidence around the SLR peaks to get the highest value of field intensity enhancement. The field distribution in the planes *xz* and *yz* through the center of the nanorod, and in the plane *xy* at a height $z = 3$ nm for 0L and $z = 8$ nm for 16L above the topmost surface of the particle array are shown in Figures 5c, d, and e, respectively.

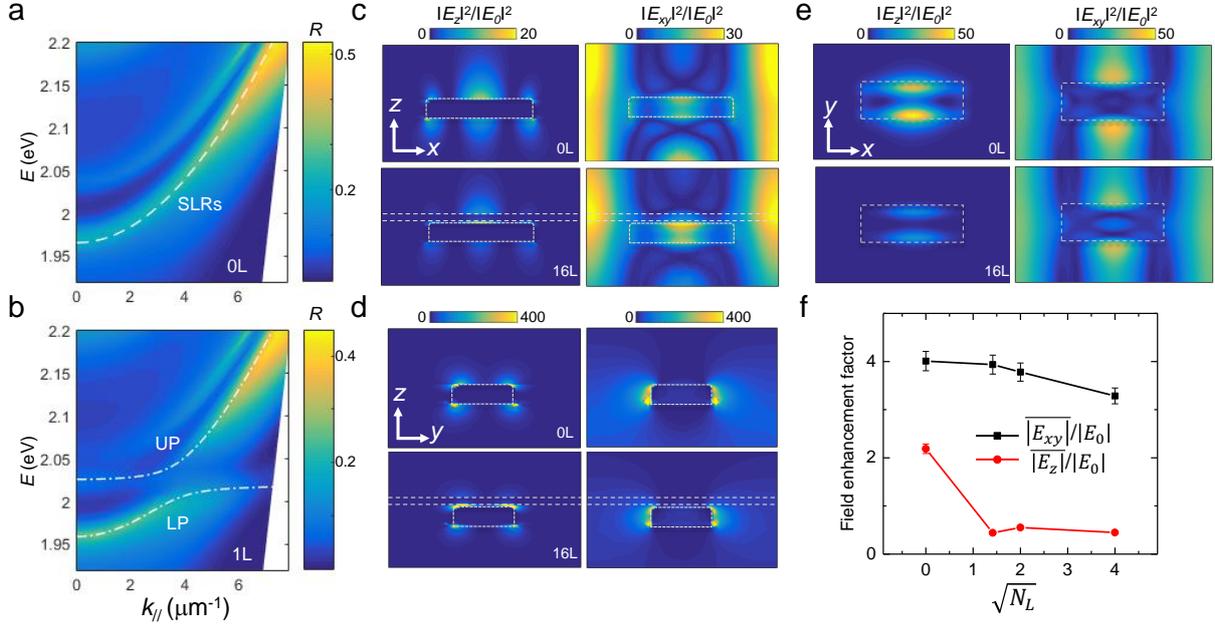

**Figure 5.** *Finite-element simulations of spectra and field distributions*. Angle-resolved reflection spectra simulation of the bare particle array (a), and of the 1L WS$_2$ on the array (b). The spectra are simulated for *p*-(TM) polarization and the angular dispersion of the 1L on the array is fitted with the coupled oscillator model. The dashed-white curve in (a) describes the dispersion of the SLRs of the bare particle array. The upper and lower dashed-dotted white curves in (b) represent the dispersion of the UP and LP band, respectively. Simulated out-of-plane ($|E_z|^2/|E_0|^2$, left column) and in-plane ($|E_{xy}|^2/|E_0|^2$, right column) field intensity enhancement at the energy of A excitons in the planes xz (c), yz (d), and xy (e) of a lattice unit cell. In (c) and (d), the field is analyzed by cutting through the center of the nanoparticle. The top and bottom row of the images in (c)-(e) corresponds to the result of 0L and 16L on the particle array, respectively. In (e), the field is calculated at a height of $z = 3$ nm above the array of the bare SLRs and $z = 8$ nm for 16L on top of the array. The dashed rectangular and horizontal lines in the field distribution images (c)-(e) represent the boundaries of the nanoparticles and WS$_2$ flake, respectively. (f) Field amplitude enhancement factor as a function of the square root of $N_L$. The field is averaged over a lattice unit cell at the height corresponding to the center of the flake. The black squares and red circles represent the in-plane and out-of-plane field components, respectively.

Due to the high permittivity contrast between WS$_2$ and its dielectric surrounding, the normal field ($|E_z|^2/|E_0|^2$ shown in the left column of Figures 5c and d) displays a strong discontinuity and is largely screened in the high refractive index WS$_2$ layer, as a consequence of the boundary condition of Maxwell equation (continuity of $\varepsilon E_z$). In contrast, the profile of the in-plane ($|E_{xy}|^2/|E_0|^2$) field is only slightly modified by the existence of the WS$_2$ flake (right column of Figures 5c and d) except for a slightly weaker intensity. The in-plane electric field is not only confined to the particles but also extends into the surrounding dielectric medium. However, the out-of-plane component is mostly localized close to the particles. As shown in Figure 5f and Figure S11, both absolute amplitude enhancement factors, averaged over a lattice unit cell, decrease with WS$_2$ on top of the particle array and the in-plane field is ~ 8 times larger than the out-of-plane component. The different spatial overlap between in-plane and out-of-plane field components, and the screened out-of-plane field in WS$_2$ suggest that the interaction of excitonic dipoles with the normal field can be disregarded. Therefore, we can calculate the ratio of coupling efficiency of SLRs with ML and 1L based on the relationship, $\Omega_R \propto 2\sqrt{N_L} E \cdot \mu \approx 2\sqrt{N_L} |E_{xy}| \mu_{xy}$, with in-plane excitonic dipole $\mu_{xy}$ (Figure 1d) and field distribution in the flake $|E_{xy}|$ (Figure 5f). For simplicity in this calculation, we have used the averaged field amplitude, $\overline{|E_{xy}|}$, in a single lattice unit. We plot the estimated Rabi energy of ML as blue triangles in Figure 4c, finding a remarkable good agreement with the measurements. This result confirms that the Rabi energy in our system stems mostly from in-plane coupling and saturates around 16L. Consequently, the periodic nanoparticle array can essentially strongly couple with both 1L and ML.

Using $\Omega_R = 2\sqrt{N} \boldsymbol{E} \cdot \boldsymbol{\mu}$, we can give an estimate of the number of coupled excitons, $N$, in WS$_2$ using the dipole moments of WS$_2$ and the vacuum field amplitude $E = \sqrt{\hbar\omega/(2\varepsilon\varepsilon_0 V)}$.[18,52] The mode volume ($V$) of delocalized SLRs can be defined by the in-plane coherent scattering length and the out-of-plane decay length of the field. We estimate the coherence length as ~ 1.7 μm from the group velocity of SLRs ($v_g$ ~ 0.08 c) at the resonant wave vector obtained from the dispersion and dephasing lifetime ($\tau$ ~ 68 fs) obtained from the resonance linewidth.[53] Hence, $E$ supported by the particle array is estimated to ~ 3.5 x 10$^5$ V/m and $N$ is ~ 3000 for the case of 1L. This value of $N$ is the of same order as reported by the coupling of 1L WS$_2$ with propagating surface plasmon polaritons on a metallic grating.[54]

In summary, we have demonstrated the strong coupling of direct transition excitons in monolayers and multilayers of WS$_2$ with collective plasmonic resonances in open plasmonic cavities formed by arrays of metallic nanoparticles at room temperature. The Rabi energy increases by increasing the number of layers. However, this increase does not scale with the square root of the number of layers. Only in-plane coupling contributes to the Rabi energy on the nanoparticle array due to the screened out-of-plane field distribution in WS$_2$. The decreased in-plane dipole moments of thicker WS$_2$ multilayers result in a smaller Rabi energy than the expected value from collective strong coupling. Understanding the mechanisms that increase and limit the coupling strength in 2D semiconductors and controlling this coupling is of utmost importance for the modification of the material properties by light-matter hybridization and the development of polaritonic devices. For instance, exciton-polaritonic states with ultrashort lifetimes in TMD multilayer systems could be used to suppress the inter-valley scattering and consequently enhance the degree of valley/spin polarization at room temperature. Polaritonic states could be also used to enhance the photoluminescence quantum yield of the direct transition excitons in multilayers. Furthermore, the collective and delocalized properties of polaritons and the investigated open cavities of nanoparticle arrays could be promising platforms to enhance exciton transport,[29,30] and to achieve long-distance energy transfer in 2D semiconductor devices.[31,32]

**METHODS**

**Sample fabrication.** The silver particle array was fabricated onto a fused silica substrate by electron beam lithography and passivated with a 3 nm thick layer of Al$_2$O$_3$ deposited with atomic layer deposition at ~ 100 °C. High-quality atomically thin WS$_2$ flakes were mechanically exfoliated from a synthetic single crystal (hq graphene), and the thicknesses of the flake samples were determined with white light microscope imaging, extinction, PL spectroscopy and atomic force microscopy (AFM). The $WS_2$ flakes were exfoliated onto an optically transparent and flexible PDMS substrate. The flakes on the PDMS were aligned and softly deposited under a microscope onto the silver particle array by surface adhesion. The thickness of the 16 layers of $WS_2$ was measured with AFM to be 9.9 ± 0.4 nm (Figures S1 and S2).

**Transmission, reflection and PL measurements.** The transmittance spectra of the sample were measured using an optical microscope under normal incidence. The samples were aligned along the optical axis of the microscope and illuminated with quasi-collimated white light from a halogen lamp. The light transmitted by the samples was collected using a microscope objective lens (Nikon CFI S Plan Fluor ELWD 60x, NA = 0.7) and imaged by a spectrometer (Princeton Instruments SpectraPro 300i) and an electron-multiplying charge-coupled device camera (Princeton Instruments ProEM: 512). The extinction is recorded as 1-T, where T is the transmittance of the sample. The reflection and PL spectra of all the samples were measured using a microscope reflectometry setup equipped for optical Fourier analysis. The setup is described in detail elsewhere.[15,22]

**Angular dispersion of SLRs.** The SLR dispersion curves extracted from the peaks of the reflection spectra of the bare particle array were fitted with a two coupled oscillator model, describing the interaction between the LSPRs of the individual particles and degenerate (0, ±1) RA respectively.[39] The RA dispersion follows the grating equation $\pm k_{//d} = k_{//i} + G$, where $k_{//d} = (2\pi/\lambda)n_{eff} \cdot \mathbf{u_d}$ and $k_{//i} = (2\pi/\lambda)\sin(\theta) \cdot \mathbf{u_i}$ are the parallel-to-the-surface components of the diffracted and incident wave vectors, respectively, $\mathbf{u_d}$ and $\mathbf{u_i}$ are the unitary vectors along the diffracted and incident directions projected on the plane of the array, $G = (2\pi m/p_x, 2\pi n/p_y)$ is the reciprocal lattice vector with $(m, n)$ defining the diffraction order and $(p_x, p_y)$ the lattice constants of the array, $\theta$ is the angle between the wave vector of the incident beam and the direction normal to the plane of the array, and $n_{eff}$ is the effective refractive index defining the phase velocity of the RAs.

**Numerical simulations.** The simulations were performed using the commercial finite-element method software COMSOL Multiphysics 5.2. We simulate only one unit cell of the periodic array by surrounding a single silver nanorod with Floquet periodic conditions. The nanorods were modeled as parallelepipeds where we replaced the sharp corners by sphere sections of radius 10 nm to avoid extreme hotspots and to approach the experimental conditions. A dispersionless model was used for the fused silica substrate, alumina spacer and PDMS superstrate. The optical constants of silver were obtained from Palik's handbook,[55] while the optical constants of $WS_2$ are obtained from the experimental results shown in Figures S4c and d. Considering the little difference between in-plane and out-of-plane dielectric constant of $WS_2$,[56] we used the uniform background refractive index ($n = 2.76$) for $WS_2$ crystal during the simulation. To avoid the problem of meshing layers smaller than the nanoscale, we simulate the interaction with the $WS_2$ monolayer with a surface conductivity, related to the permittivity as $\sigma = -i\epsilon_0 \omega d [\varepsilon_{WS_2}(\omega) - 1]$,[57] with $d$ the thickness of the $WS_2$ monolayer and $\varepsilon_{WS_2}$ its permittivity.

## ASSOCIATED CONTENT

**Supporting Information**

The description of sample I and II, refractive index and absorption spectra of $WS_2$, angle-resolved reflection spectra of the bare flake, coupled oscillator model fits to sample I and II, normal incidence spectra of the flake on array, and elctric field enhancement factor are attached in the supporting information (SI).

## AUTHOR INFORMATION

**Corresponding Author**

*E-mail: s.wang@differ.nl18

*E-mail: j.gomez.rivas@tue.nl

**Notes**

The authors declare no competing financial interest.

## ACKNOWLEDGEMENTS

We thank Dr. Marcos H. D. Guimarães for measuring the thickness of the sample by AFM. This work was financially supported by the Netherlands Organisation for Scientific Research (NWO) through the Industrial Partnership Program "Nanophotonics for Solid-State Lighting" between Philips and NWO through the Innovational Research Activities Scheme (Vici project number 680-47-628). A.G.C. was supported by a Marie Curie International Outgoing Fellowship. F.V. and B.M. acknowledge support from Le Fonds pour la Formation à la Recherche dans l'Industrie et dans l'Agriculture (FNRS-FRIA) in Belgium.

# Supplementary Information

# Screened Strong Coupling of Excitons in Multilayer $WS_2$ with Collective Plasmonic Resonances

**Shaojun Wang,**[†*] **Quynh Le-Van**,[†] **Fabio Vaianella,**[#] **Bjorn Maes**,[#] **Simone Eizagirre Barker,**[†] **Rasmus H. Godiksen,**[‡] **Alberto G. Curto,**[‡] **and Jaime Gomez Rivas**[†,‡*]

[†] Dutch Institute for Fundamental Energy Research, Eindhoven, The Netherlands

[#] Micro- and Nanophotonic Materials Group, Faculty of Science, University of Mons, 20 place du Parc, B-7000 Mons, Belgium

[‡] Department of Applied Physics and Institute for Photonic Integration, Eindhoven University of Technology, Eindhoven, The Netherlands

[*]E-mail: s.wang@differ.nl
[*]E-mail: j.gomez.rivas@tue.nl

1. Sample I
2. Sample II
3. Refractive index and absorption spectra of $WS_2$
4. Angle-resolved reflection spectra of the bare flake
5. Coupled oscillator model fits to sample I
6. Coupled oscillator model fits to sample II
7. Normal incidence spectra of the flake on array
8. Electric field enhancement factor

1. **Sample I**

The WS$_2$ flake is sandwiched between the soft adhesive PDMS superstrate and the nanoparticle array on the quartz substrate. Hence, peeling off the PDMS layer on top to measure the thickness of the flake could easily break it. In order to determine the thickness of sample I, we chose to use another WS$_2$ flake, also sandwiched between PDMS and quartz, as a reference. This flake shows similar optical contrast with the ML region of sample I under the inverted microscope. The average transmittance of two multilayer regions on the reference sample (ML1 and ML2, encompassed with green dotted curves in Figure S1a) is slightly different to the transmittance of sample I (Figure S1c), which indicates that the thickness of the sample I is between the thickness of ML1 and ML2. After peeling off the PDMS on the reference sample, we measured its thickness by AFM. The AFM image of the reference flake and the height profile along the white dashed line on ML1 are shown in Figure S1b and d, respectively. The thicknesses of ML1 and ML2 are 8.3 ± 0.4 nm (i.e., 13 layers) and 11.4 ± 0.4 nm (18 layers), respectively. Accordingly, the thickness of the target ML sample is determined to be 9.9 ± 0.4 nm, which corresponds to 16 WS$_2$ monolayers using the layer distance of 0.62 nm.[1,2]

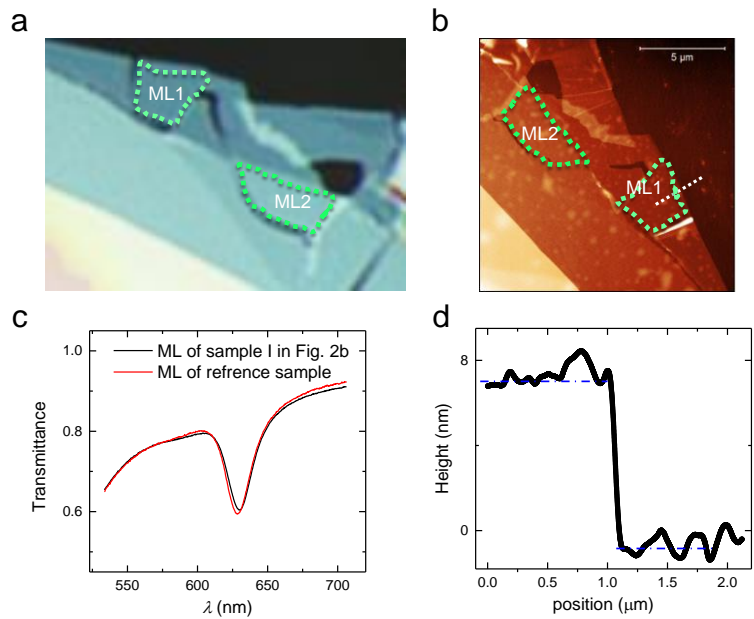

**Figure S1 (a)** White light image of the reference WS$_2$ flake for measuring the thickness of the ML on sample I. The optical contrast of the multilayer regions (ML1 and ML2) in this flake are similar to the ML area of the sample shown in Fig. 2b, which is confirmed by the microscope transmission spectra in (c): the averaged transmittance spectra of ML1 and ML2 (red solid line) almost overlaps with that of the ML sample used in the main text. **(b)** Atomic force microscope (AFM) image of the reference WS$_2$ flake. The height profile along the white dashed line on ML1 in (b) is shown in (d).

We have also directly measured the thickness of sample I by peeling off the PDMS superstrate after the optical measurements. From the white light and PL images of the sample before (Figure S2a and b) and after (Figure S2c and d) peeling off the PDMS, we notice that most regions of the flake remain on the quartz substrate except the monolayer region on the particle array. The removal of the monolayer by peeling off the PDMS is the reason why we initially used the reference sample to characterize the thickness of the sample. The AFM image (Figure S2e) of the ML region of interest in sample I (white square in Figure S2c) gives a thickness of 10.1 ± 0.7 nm (Figure S2f) in agreement with the previous measurements.

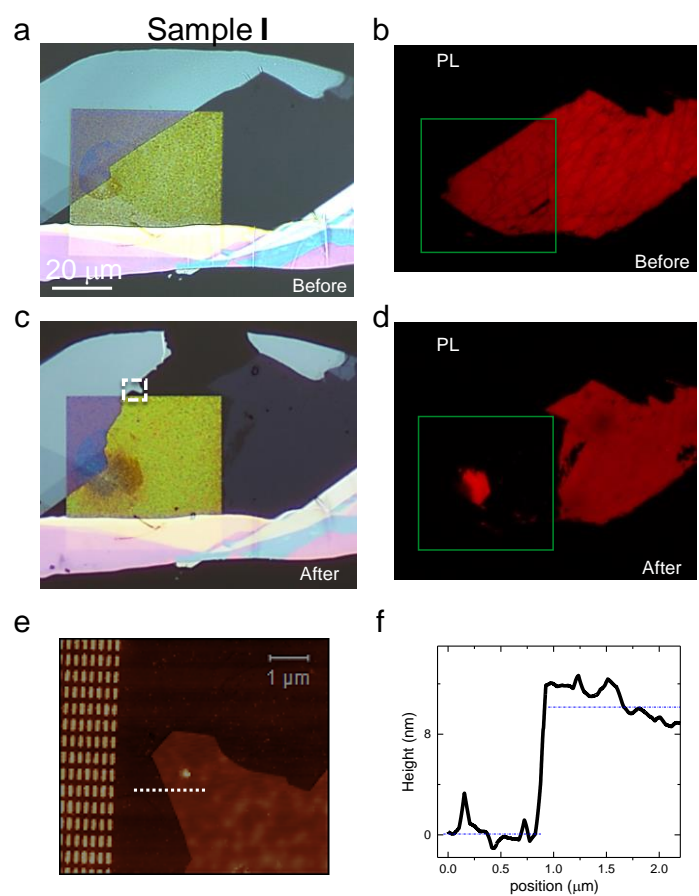

**Figure S2** (a)(c)White light and (b)(d) PL images of sample I before (a)(b) and after (c)(d) peeling off the PDMS superstrate. (e) AFM image of the sample region marked by the square in (c). The height profile along the white dashed line in (e) is shown in (f).

2. **Sample II**

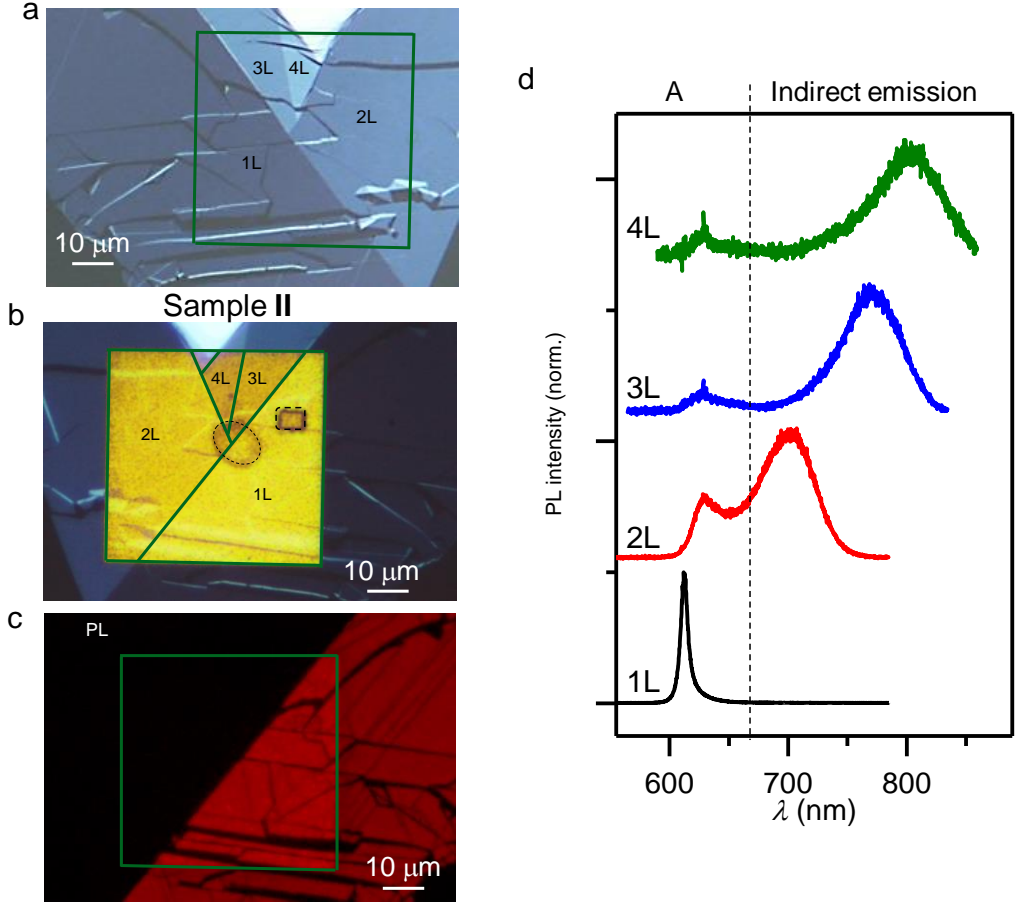

**Figure S3** (a)(b)White light image of sample II before and after transfer onto the particle array. (c) PL image of sample II after transfer onto the array. The green square indicates the position of the particle array. The bilayer (2L), trilayer (3L), and four-layer (4L) regions are confirmed from the peak wavelengths of indirect band gap emission shown in the PL spectra of (d). These values are comparable with those in the literature.[3,4]

3. **Refractive index and absorption spectra of WS$_2$**

In order to calculate the in-plane component of the dielectric permittivity, the transmission spectra of the WS$_2$ (Figures S4a and b) are fitted using the transfer matrix method (TMM) and a multi-Lorentizan model:[1,5,6]

$$\varepsilon(E) = \varepsilon_B + \sum_{j=1}^{N} \frac{f_j}{E_{0j}^2 - E^2 - iE\gamma_j}, \quad (1)$$

where $E$, $\varepsilon_B$, $f_j$, $E_{0j}$ and $\gamma_j$ are respectively the photon energy, the non-resonant background dielectric permittivity, oscillator strength, resonance center energy, and the linewidth of the absorption band $j$. The absorptance (Figure 1c) was determined using the TMM and considering the conservation of energy, i.e., $A=1-T-R$, where $T$ is the transmittance of the normal incident beam through the samples and $R$ is the reflectance. Figure S5a shows the pure absorptance of A excitons. The corresponding linewidth and absorptance per layer at peak position are summarized in Figure S5b.

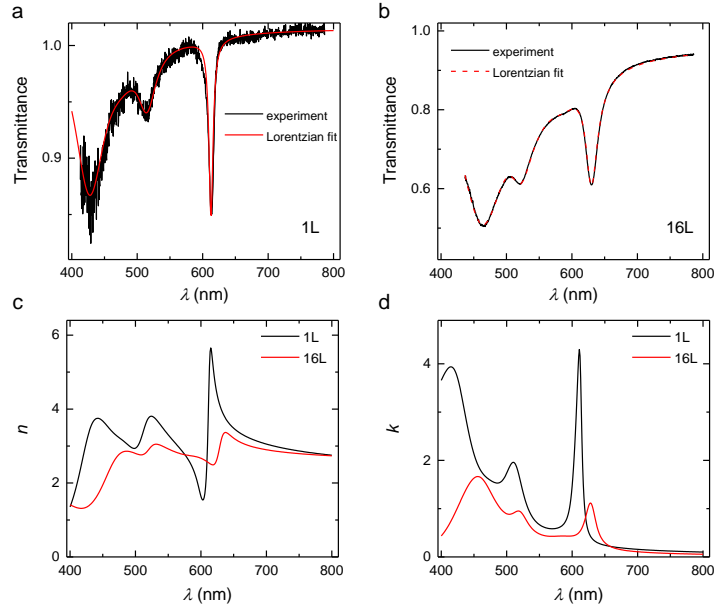

**Figure S4** Transmission spectra of the 1L in sample II (a) and the 16L in sample I (b) characterized under normal incidence, and fitted with a multi-Lorentzian model. Real (c) and imaginary (d) components of the in-plane refractive index of 1L and 16L $WS_2$.

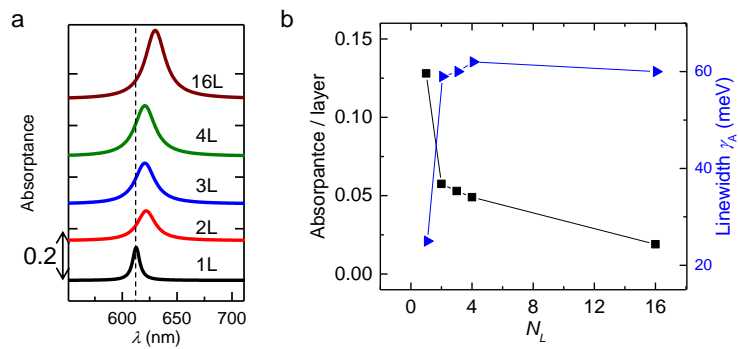

**Figure S5** (a) Normal incidence absorption of the A excitons in 1L to 4L $WS_2$ in sample II, and 16L $WS_2$ in sample I. (b) The linewidth and absorptance per layer at the peak position of each region extracted from (a). The solid lines in (b) are guides to the eye.

## 4. Angle-resolved reflection spectra of the bare flake

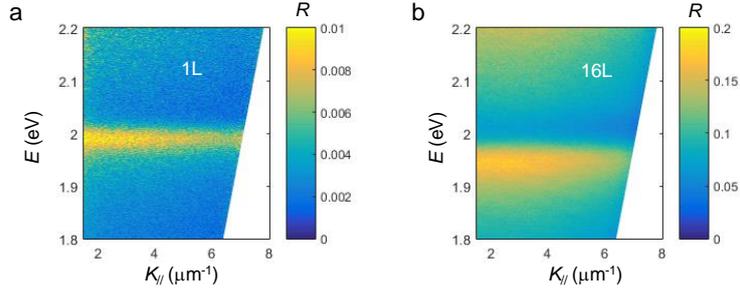

**Figure S6** Angle-resolved reflection spectra of the bare 1L (a) and 16L WS$_2$ (b) on the quartz substrate of sample I, analyzed in TM polarization.

## 5. Coupled oscillator model fits to sample I

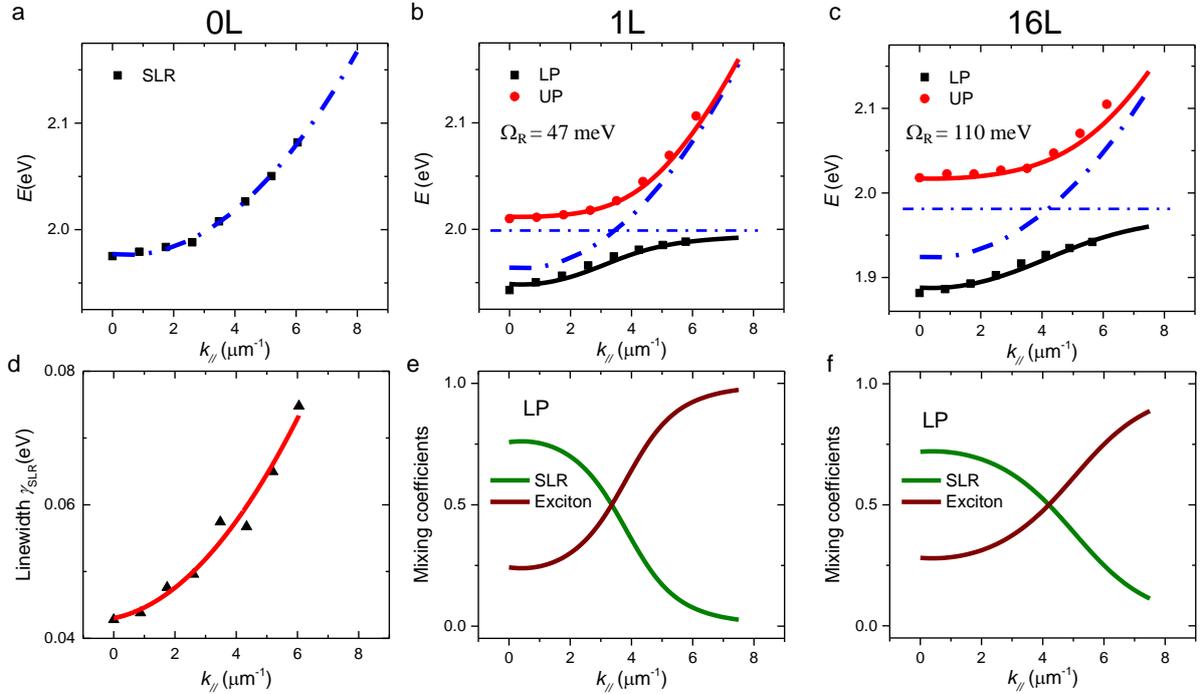

**Figure S7** (a) Energy and (d) linewidth of bare SLRs extracted from Fig. 3(d). (b)(c) Lower polariton (LP, black squares and curve) and upper polariton (UP, red dots and curve) bands extracted from Figs.3 (e) (f) and fitted by the two coupled oscillator model. The blue horizontal dash-dotted line and parabolic curve represent the A excitonic resonance and bare SLRs, respectively. (e)(f) Weight fractions, i.e., mixing coefficients, of the LP as a function of $k_{//}$.

## 6. Coupled oscillator model fits to sample II

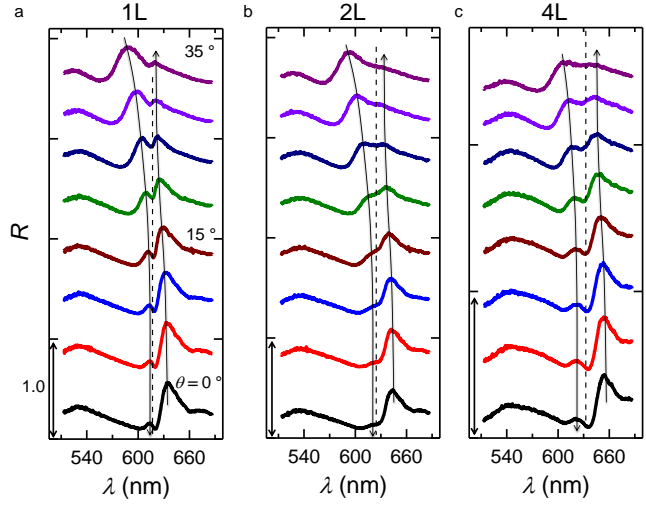

**Figure S8** Angle-resolved reflection spectra of the 1L (a), 2L (b) and 4L WS$_2$ (c) of sample II on the particle array, analyzed in TM polarization. The vertical dashed line indicates the central wavelength of bare excitonic resonance. The spectra are measured from $\theta = 0°$ to $35°$ from the bottom to the top.

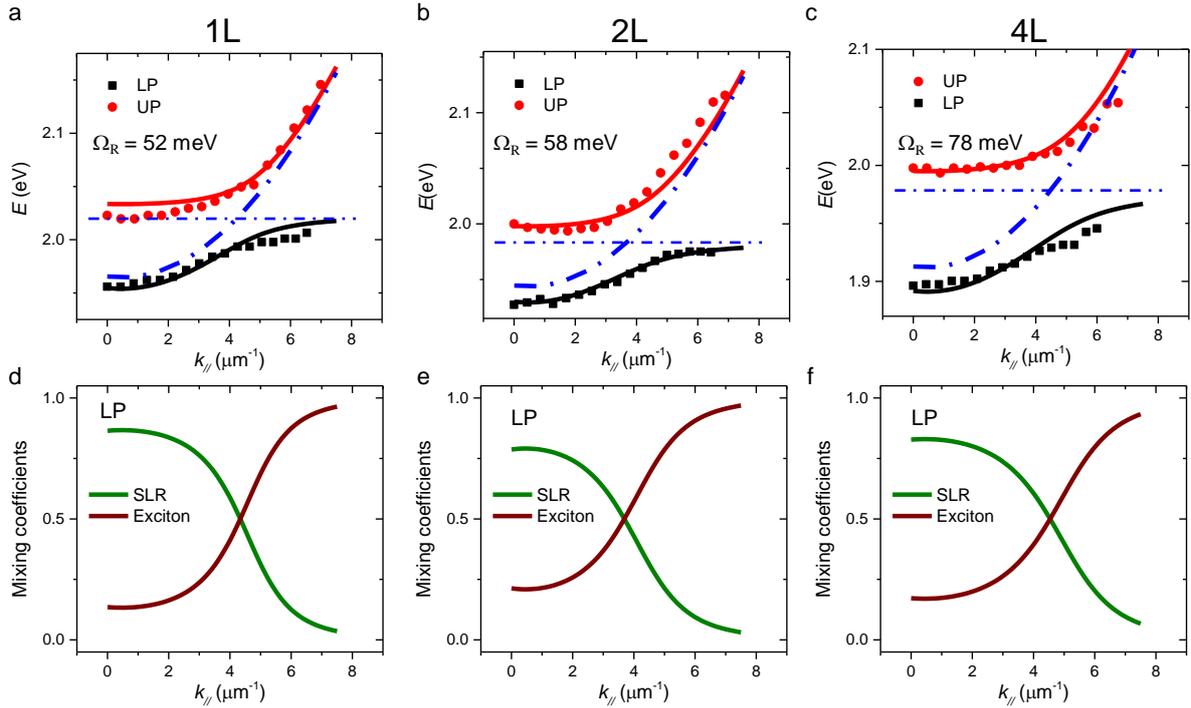

**Figure S9** (a)-(c) Lower polariton (LP, black squares and curve) and upper polariton (UP, red dots and curve) bands extracted from the Fig. S8 and fitted by the two coupled oscillator model. The blue horizontal dash-dotted line and parabolic curve represent the A excitonic resonance and bare SLRs, respectively. (d)-(f) Weight fractions, i.e., mixing coefficients, of the LP as a function of $k_{//}$.

## 7. Normal incidence spectra of the flake on array

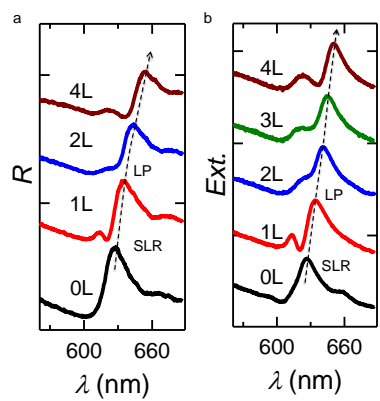

**Figure S10** Comparison of the reflection (a) and extinction spectra (b) of the particle array with different thickness of flake on top and measured at normal incidence.

## 8. Electric field enhancement factor

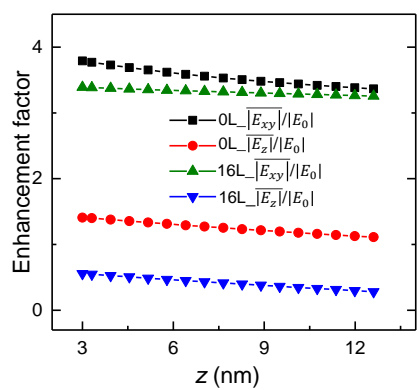

**Figure S11** Averaged electric field amplitude enhancement factor of bare array (0L: black squares for the in-plane component, red dots for the out-of-plane component) and 16L on the array (green triangles for the in-plane component, blue triangles for the out-of-plane component) as a function of the height $z$ above the array. The enhancement factor is calculated from z = 3 to 13 nm (with a $\Delta z = 0.5$ nm), which corresponds to the thickness of the 16L $WS_2$.

**Supplementary References**